\documentclass{elsart}
\usepackage{color}
\usepackage{graphicx}
\usepackage{epsfig}
\usepackage{amssymb}

\begin{document}
\begin{frontmatter}

\title{Development of a thermal ionizer as ion catcher}
\author{E. Traykov\corauthref{cor1}}{, }
\ead{traykov@kvi.nl}
\author{U. Dammalapati}{, }
\author{S. De}{, }
\author{O.C. Dermois}{,\,}
\author{L. Huisman}{,\,}
\author{K. Jungmann}{,\,}
\author{W. Kruithof}{,\,}
\author{A.J. Mol}{,\,}
\author{C.J.G. Onderwater}{,\,}
\author{A. Rogachevskiy}{,\,}
\author{M. da Silva e Silva}{,\,}
\author{M. Sohani}{,\,}
\author{O. Versolato}{,\,}
\author{L. Willmann}{,\,}
\author{H.W. Wilschut}
\address{Kernfysisch Versneller Instituut, University of
Groningen, Zernikelaan 25, 9747 AA Groningen, The Netherlands}
\corauth[cor1]{Corresponding author. Tel.: +31 503633569, fax: +31
503633401}

\begin{abstract}
An effective ion catcher is an important part of a radioactive beam
facility that is based on in-flight production. The catcher stops
fast radioactive products and emits them as singly charged slow
ions. Current ion catchers are based on stopping in He and H$_2$
gas. However, with increasing intensity of the secondary beam the
amount of ion-electron pairs created eventually prevents the
electromagnetic extraction of the radioactive ions from the gas
cell. In contrast, such limitations are not present in thermal
ionizers used with the ISOL production technique. Therefore, at
least for alkaline and alkaline earth elements, a thermal ionizer
should then be preferred. An important use of the TRI$\mu$P facility
will be for precision measurements using atom traps. Atom trapping
is particularly possible for alkaline and alkaline earth isotopes.
The facility can produce up to 10$^9$ s$^{-1}$ of various Na
isotopes with the in-flight method. Therefore, we have built and
tested a thermal ionizer. An overview of the operation, design,
construction, and commissioning of the thermal ionizer for TRI$\mu$P
will be presented along with first results for $^{20}$Na and
$^{21}$Na.
\end{abstract}
\end{frontmatter}

\section{Introduction}
\label{intro} In the TRI$\mu$P facility~\cite{tra07a} radioactive
nuclides are produced in a gas target using inverse reaction
kinematics~\cite{tra07b} and separated from the beam in a magnetic
separator~\cite{ber06}. Typical energies range up to several 10 MeV
per nucleon. The secondary beams from the magnetic separator have
large transverse emittances and wide energy distributions.
Thermalization is required in order to efficiently collect the
produced isotopes and transport them further as a beam of ions.

A gas stopper~\cite{arj85} suitable for slowing of a large range
of isotopes was initially considered as an ion catcher for the
TRI$\mu$P facility, but reduction of the extraction efficiency at
high intensities~\cite{huy02}, neutralization limits~\cite{wil06},
and the required gas purity~\cite{den06} led to the decision to
use an alternative slowing method. This method employs a thermal
ionizer (TI) ion catcher, the operation of which is based on hot
cavity ion sources~\cite{kir81} such as used in ISOL (Isotope
Separation On-Line) facilities. A TI allows high efficiencies to
be reached  for short-lived alkali and alkali earth isotopes which
are of primary interest for the goals of TRI$\mu$P~\cite{wil99,
jun02, wil03, jun05}. Moreover, the efficiency of the TI is
independent of the secondary beam intensity.

In contrast to ISOL systems where the primary beam is fully stopped
in the target-ion source system, at the TRI$\mu$P facility only
secondary isotopes (with several orders of magnitude lower intensity
than the primary beam) are implanted in the thermal ionizer and thus
activation of the ionizer material is negligible. This allows
maintenance of the ionizer without radiation safety issues involved.
We have made use of this in optimizing our initial design. This
feature can also be used to accelerate the development of ISOL
sources.

The operation of the TI is based on stopping in thin foils,
diffusion to the foils' surface, adsorption/desorption on the TI
walls and foils, ionization/neutralization, and extraction with
electric fields (see Fig.~\ref{fig:sim}).

\subsection{Beam spot and stopping range}
The next step after production and separation in the magnetic
separator is to ensure that the isotopes of interest will be
implanted in the foils of the Thermal Ionizer. For this the TI is
placed on the optical axis in the focal plane of the magnetic
separator where the beam spot has minimal lateral size (smaller than
the dimensions of the stopper foils which have 2.5 cm diameter). To
ensure that the ions of interest are implanted in the stopper
material the total thickness of the stopping material is chosen by
taking into account the maximal energy acceptance of the separator
($\pm$ 4 $\%$) and the thickness inhomogeneity of the materials
(typically below 10 $\%$). The necessary stopping material thickness
is estimated with the program SRIM~\cite{zie85} using the mean
energy of the produced nuclides which is determined by the magnetic
rigidity settings of the separator. The implantation depth of the
ions in the foils stack can be optimized by using a rotatable Al
degrader upstream of the TI.

\subsection{Diffusion}
The implanted isotopes inside the Thermal Ionizer foils come to the
surface by diffusion. Steady state diffusion is governed by Fick's
first law~\cite{fic55}
\begin{equation}
\vec{j}=-D\nabla{C}(\vec{r}), \label{eq:TIdiffeq2}
\end{equation}
where $C$ is the concentration of the diffusing particles per unit
volume at a given position $\vec{r}$ in the foil, $\vec{j}$ is the
flux of the particles at any position $\vec{r}$ inside the foil
volume, and $D$ is the diffusion constant. For a given element and
material combination $D$ depends on the material temperature $T$
\begin{equation}
D(T)=D_0e^{-\frac{E_0}{RT}}, \label{eq:TIdiffconst}
\end{equation}
where $D_0$ and $E_0$ are the Arrhenius coefficients for diffusion,
representing the maximum diffusion constant $D$ for
$T\rightarrow\infty$ and the activation energy for diffusion,
respectively.

Analytical solutions of the diffusion equation~\cite{cra79, fuj81}
for a foil with thickness $d$ lead to a fractional release
efficiency as a function of time $t$ (assuming homogeneous
deposition of the isotopes and desorption from both surfaces)
\begin{equation}
\varepsilon_d(t)=
2{\sqrt{\frac{t{D}}{{d^2}}}}{\tanh\left({\sqrt{\frac{{d^2}}{4t{D}}}}\right)}.
\label{eq:TIdiffreleaase}
\end{equation}
This allows an estimate of efficiencies for radioactive isotopes by
substituting $t$ with the half-life of the radioactive isotope
$\tau_{1/2}$. The release efficiencies for $^{20}$Na
($\tau_{1/2}=0.448$ s) and $^{21}$Na ($\tau_{1/2}=22.5$ s) from a 1
$\mu$m thick foil were calculated for various values of the
diffusion constant $D(T)$~\cite{tra06}. The dependence shows that
for diffusion constant values below 10$^{-14}$ m$^2$/s up to 7.1
times higher release efficiency is expected for $^{21}$Na than for
$^{20}$Na independently of the foils thickness.

Measured data for $D(T)$, $D_0$, and $E_0$ are available for a large
variety of elements and materials, but diffusion data for sodium in
tungsten could not be found in the literature. However, data exist
for lithium and potassium in tungsten \cite{mcc60, lov63, bay83,
kir92}. Rough estimates for diffusion of sodium in tungsten yield
 10$^{-16}$ m$^2$/s $<D(T)<$ 10$^{-12}$ m$^2$/s for
temperatures in the TI operation region, i.e. 2500 K.

\subsection{Effusion}
After diffusion to the surface of the stopper foils the elements
need to be transported to the TI exit aperture. The extraction of
ions from the cavity through the aperture is called effusion. The
particles undergo multiple collisions with stopping foils and cavity
walls. Further there are plasma interactions which depend on the
mean free path inside the TI cavity, thus on the background
pressure.

The mean delay time for effusion $\tau_e$ is defined as the sum of
the total time spent between collisions with walls and the total
time the ions spent on the walls due to adsorption
\begin{equation}
\tau_e=\chi({\tau_f}+{\tau_a}). \label{eq:TIefftime}
\end{equation}
Here $\tau_f$ is the mean time between collisions with the walls,
$\tau_a$ is the mean ``sticking time", and $\chi$ is the mean number
of collisions before leaving the exit aperture.

Effusion delay times were estimated for $^{20}$Na and $^{21}$Na in a
TI cavity at $T=2000$ K. For $\chi$ = 2000 the accumulated surface
sticking time $\chi\tau_a$ is less than 1 $\mu$s (for Na ions on
tungsten surface at $T=2000$ K the mean sticking time is
$\tau_a=2.1\cdot10^{-10}$ s~\cite{ros84}). This process is several
orders of magnitude faster than the total time-of-flight
$\chi\tau_f$ which is of the order of a few ms. Both $\tau_f$ and
$\tau_a$ decrease with increasing temperature.

Due to the very low adsorption enthalpy ($\Delta{H}_{ads}=127.1$
kJ/mol) for sodium on tungsten the effusion related delay is several
orders smaller than the delay from the diffusion in the stopping
foils (the literature upper limit of $D(T)=10^{-12}$ m$^2$/s corresponds to at least 50 ms of diffusion delay for 1 $\mu$m thick foils). Therefore, effusion delay can be neglected for sodium extraction from the thermal ionizer.
\begin{figure}
\centerline{\includegraphics[width=12cm,angle=0,clip=true]{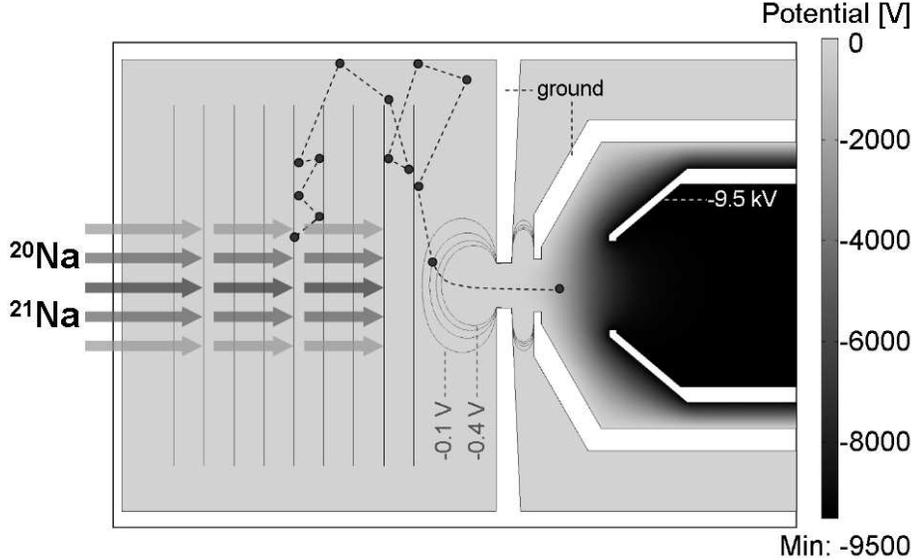}}
\caption {Electrostatic simulation of the TI. The sodium isotopes
enter the TI and stop in a stack of thin W foils. High temperature
allows fast diffusion of the elements out of the foils. Multiple
wall collisions lead to consecutive neutralization and ionization.
Penetrating electrostatic fields from an extraction electrode (-9.5
kV) increase the probability for ion extraction.} \label{fig:sim}
\end{figure}

\subsection{Ionization and electrostatic extraction}
Ionization inside the thermal ionizer is essential for the
extraction of ion beam which can be achieved by applying a negative
electric potential on an extraction electrode (Fig.~\ref{fig:sim}).

Surface ionization in collisions with the foils and the ionizer
walls is the dominant process. Charge-changing processes in the
ionizer volume, i.e. not at the surfaces, can be neglected. Surface
ionization is described by the Langmuir equation~\cite{lan25}
\begin{equation}
\alpha_s=\frac{n_{is}}{n_{as}}=\frac{g_i}{g_0}\textnormal{exp}\left(\frac{\varphi-W_i}{kT}\right),
\label{eq:TIsurfionprob}
\end{equation}
where $\alpha_s$ is the ionized fraction, $n_{is}$ and $n_{as}$ are
the ion and atom densities, respectively, near the cavity surface.
They can be considered uniform in the whole cavity volume at low
temperatures. ${g_i}$ and ${g_0}$ are the statistical weights of the
ionic and atomic states, according to the total angular momentum of
the states $J_{i,0}$ (${g_i}$/${g_0}$=1/2 for alkali elements).
$\varphi$ and $W_i$ are the work function of the ionizing material
and the first ionization potential of the element of interest. The
surface ionization efficiency is
\begin{equation}
\varepsilon_{si}=\frac{n_{is}}{n_{is}+n_{as}}=\frac{\alpha_s}{1+\alpha_s}.
\label{eq:TIsurfioneff}
\end{equation}
The ionization efficiency for sodium on a tungsten surface is
calculated~\cite{tra06} to be between 2 and 9 $\%$ for $T$ between
2000 and 3000 K. This also accounts for the temperature dependence
of the tungsten work function $\varphi$. The surface ionization
efficiency for sodium on tungsten has been measured~\cite{dat56}. It
increases from 6 to 8 $\%$ in the temperature range from 2200 to
2850 K.

The extraction efficiency $\varepsilon_{extr}$ is higher than the ionization efficiency $\varepsilon_{i}$ due to the
presence of a penetrating electrostatic field through the exit
aperture of the cavity. The field allows for ions to be extracted
when passing close to the aperture where the electric potential is
higher than the ion kinetic energy (as shown in Fig.~\ref{fig:sim}).
Meanwhile, the electrostatic extraction has no effect on neutral
particles allowing them to stay inside the cavity longer. Monte
Carlo simulations were performed to estimate the effect of the
electrostatic field inside the TI cavity. Additionally, several
different exit aperture designs were used in the simulations for
comparison. In all cases $\varepsilon_{extr}$, i.e. the charged fraction of all particles passing the extraction aperture, was between 55 $\%$ and 75 $\%$ for $\alpha_s=9\%$.

At higher temperatures (above 2400 K for tungsten~\cite{kir81}) the
total ionization efficiency is expected to increase above values for
surface ionization due to formation of quasineutral plasma inside
the cavity volume~\cite{kir90}.

\section{Technical design}
\label{des}
\begin{figure}
\centerline{\includegraphics[width=10cm,angle=0,clip=true]{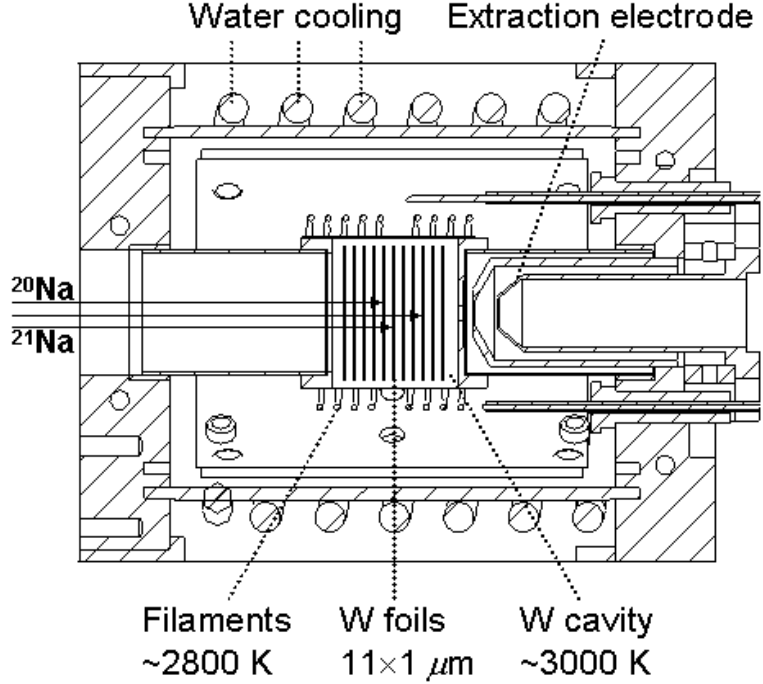}}
\caption {Schematic drawing of the Thermal Ionizer.}
\label{fig:TIsch}
\end{figure}
The material of choice for the cavity walls and stopping foils is
tungsten because of its high working temperature allowing
utilization of the hot cavity ionization~\cite{der05}. Moreover,
diffusion related delay is highly reduced at high temperatures which
is beneficial for short-lived isotopes. The tungsten cavity of the
TI is heated by electron bombardment from two tungsten filaments
outside the cavity. The temperature of the TI cavity is increased by
adjusting the power of the electron current. A stack of 11 foils of
1 $\mu$m is mounted on a tungsten frame and placed inside the TI
cavity. The setup is shown in Fig.~\ref{fig:TIsch} and the main cavity elements are listed in Table 1.
\begin{table}[ht]\label{Tab1}\caption{Mechanical specifications of the Thermal
Ionizer.}
\centering
\renewcommand{\arraystretch}{1.0}
\vspace{\baselineskip}
\begin{tabular}{ l r l }
  \hline
Cavity length               &   25     &    mm             \\
Cavity diameter             &   30     &    mm             \\
Extraction aperture diameter&   2      &    mm             \\
Stopping foil diameter      &   25     &    mm             \\
Stopping foil thickness     &   1      &    $\mu$m         \\
Distance between foils      &   1      &    mm             \\
Number of foils             &   11     &                   \\

  \hline
\end{tabular}
\end{table}

\begin{figure}
\centerline{\includegraphics[width=12cm,angle=0,clip=true]{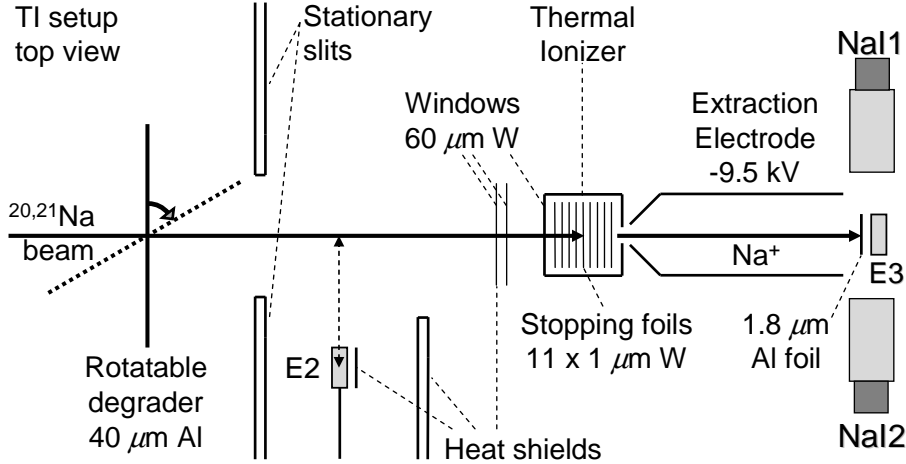}}
\caption {Schematic drawing of the Thermal Ionizer setup.}
\label{fig:setup}
\end{figure}

\section{Commissioning results}
\label{com} The isotopes employed for commissioning the thermal
ionizer were $^{20}$Na ($T_{1/2}=$ 448 ms) and $^{21}$Na ($T_{1/2}=$
22.5 s). The experimental setup is shown in Fig.~\ref{fig:setup}. To
allow varying of the implantation depth a rotatable 40 $\mu$m Al
degrader was used for fine adjustment of the energy of the incoming
ions. A thin Si detector ($E2$) upstream of the TI served for energy
loss and time-of-flight measurements for identification and
optimization of the incoming flux of secondary beam. A 1.8 $\mu$m
thick aluminium foil was placed in front of Si detector ($E3$) in
order to stop the ions extracted from the TI. $E3$ was positioned
behind the foil in order to measure the rate of delayed $\alpha$
particles from the $\beta$-decay of $^{20}$Na. The alpha lines and
branching ratios are well known~\cite{cli89}. The rate of $\alpha$
particles was measured in order to obtain the detection efficiency
of a NaI detector pair measuring electron-positron annihilation
photons. The latter was necessary for the detection of extracted
$^{21}$Na ions where no delayed $\alpha$ particles exist.

\begin{figure}
\centerline{\includegraphics[width=10cm,angle=0,clip=true]{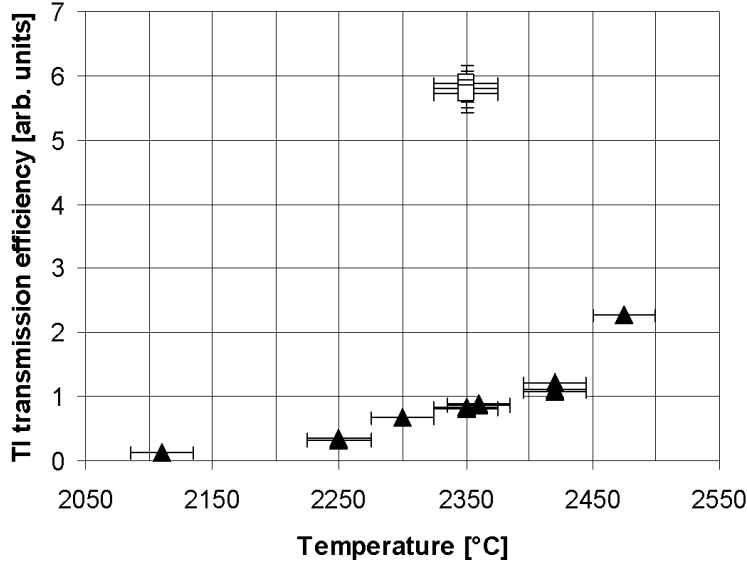}}
\caption {Measured TI efficiencies for $^{20}$Na ($\blacktriangle$)
and $^{21}$Na ($\square$) isotopes depending on the temperature of
the TI cavity. The absolute efficiency depends on various parameters
which are still being optimized. The vertical error is determined by statistics.} \label{fig:eff}
\end{figure}

The rate of the extracted $^{20}$Na ions was measured for different
ionizer temperatures (Fig.~\ref{fig:eff}). $^{20}$Na lines were
detected first at 2100 $^{\circ}$C. The rate of $^{20}$Na is
increased by approximately 25 times at 2470 $^{\circ}$C due to
decreased diffusion delay and increased ionization efficiency. The
$^{20}$Na relative efficiency is plotted in Fig.~\ref{fig:eff}
together with data for $^{21}$Na at 2350 $^{\circ}$C. We have
measured at most 48(3) $\%$ efficiency for $^{21}$Na with the current setup. The efficiency difference for $^{20}$Na and $^{21}$Na is
explained by the lifetimes of the isotopes in combination
with the diffusion-related delay in the TI. The data allows to
estimate the diffusion constant for sodium in tungsten to be at most $10^{-14}$ m$^2$/s at 2350 $^{\circ}$C.

The slowed radioactive ion beams were transported to the trapping
chamber. The beam line includes a radio frequency quadrupole (RFQ)
cooler/buncher system and a drift tube system in the low energy
beam line (LEBL).

\section{Outlook}
\label{concl} A thermal ionizer has been designed, built and
commissioned as an ion catcher for the TRI$\mu$P facility. Design
modifications were recently made in order to provide stable long
term operation at temperatures above 2500 $^{\circ}$C. The main
modification utilizes a new design for the support of the filaments. The efficiencies are expected to be significantly increased at higher temperatures due to shorter diffusion times in 0.75 $\mu$m thick
foils and the increasing significance of the hot cavity effect.
Other design modifications include changes in the design of the
extraction aperture which will allow to achieve better transmission
to the following stages of the low energy beam line of the TRI$\mu$P
facility.

The next experimental steps will include neutralization of
radioactive ions for laser trapping in a magneto-optical trap
(MOT). From the present data and the observation in~\cite{huy02} it can be concluded that a TI is a more efficient ion catcher than current gas stoppers for an in-flight production of Na isotopes with high intensities. A thermal ionizer is expected to be suitable also for slowing of other chemically similar elements.


\end{document}